\documentstyle[psfig]{l-aa}
\begin{document}
\thesaurus{08         
              (12.04.1;  
               12.07.1;  
               08.12.2;  
               08.02.3;  
               08.16.2)} 

\title{Image subtraction with non-constant kernel solutions.} 
\author{C. Alard}
\institute
{DASGAL, 61 Avenue de l'observatoire, F-75014
 Paris, France
} 
\offprints{C. Alard}
\date{Received ......; accepted ......}
\maketitle
\begin{abstract}
 It is demonstrated that non-constant kernel solution, that can fit the spatial
 variations of the kernel can be obtained with minimum computing time. The CPU
 cost required with this new extension of the image subtraction method is almost the same as for
 fitting constant kernel solution. The method is demonstrated with a serie of
 Monte-Carlo images. Differential PSF variations and differential rotation between
 the images are simulated. It is shown that this new method is able to achieve
 an optimal results even in these difficult cases. It is clear that 
 the most frequents instrumental problems are automatically corrected by the method.
 It is also demonstrated that the method does not suffer problems with under-sampling
 of the images. To conclude the method is applied to real images of crowded field.
 It is proved that much larger sub-areas of the images can be used for the fit, while
  keeping the same accuracy in the subtracted image. This result is especially 
 important in case of variables located in low density fields, like the Huchra lens.
 Many other useful applications of the method are possible for major astrophysical
 problems. For instance supernovas search should benefit the method, but other 
 applications like Cepheids surveys in other galaxies should also find some 
 advantage in using the method. Many other applications will certainly show-up,
 since variability search is a major issue in astronomy.
\end{abstract}
\section{INTRODUCTION}
 The way variability is analyzed in astronomical images has been deeply 
 influenced by the developpement of the microlensing experiments, EROS (Aubourg 1993),
 OGLE (Udalski {\it et al.} 1994), MACHO (alcock {\it et al.} 1993), DUO (Alard \& Guibert 1997) . 
 The enormous quantity of CCD images produced by
 these experiments motivated the developpement of fast and accurate
 techniques to analyze variability. The first implementations 
 of variability search were based on an analysis of catalogues
 of stellar objects. The production of the catalogues has
 been almost exclusively done using the DoPHOT software (Schechter \& Mateo 1993).
 This type of analysis is still used in microlensing experiments, and proved
 very successful. However since we are concerned only with differential
 photometry, such data processing fails to be optimal. Purely differential
 methods like image subtraction seems to be better suited for study of
 variability in astronomical images. However the developpement of image
 subtraction technique has been long delayed by the inherent difficulty
 and complexity of the method. The first successful implementation of the 
 method has been performed by Tomaney \& Crotts (1996). Further progress
 have been made by the creation of fast optimal image subtraction (OIS) method
 (Alard \& Lupton 1998). Due to its ability to solve the full least-square
 problem, the OIS method proved to produce subtracted
 images and light curves with an accuracy approaching closely the photon
 noise limit. The OIS method is very efficient for crowded stellar fields,
 like those encountered in microlensing experiments, and could readily
 be used for massive processing of microlensing data. The current 
 implementation of the OIS works by dividing the field into small sub-areas
 where constant kernel solution are derived. In dense crowded fields, sub-areas
 as small as 128 $\times$ 128 pixels can be used, at this scale 
 kernel variations can be ignored to a good approximation. However in case 
 of very bad optics or of less dense fields, the constant kernel approximation 
 does not hold any more. For instance, in high latitude fields taken 
 for supernovae search, one does not encountered a sufficient number
 of bright objects to take an area small enough to ignore the kernel 
 variations.
 We are then naturally conducted to try to make a self
 consistent fit of the kernel variations, as described in Alard \& Lupton 
 (1998). Although, it is important to notice that even if the spatial 
 variations are fitted to order 1 only, the cost to build the least-square
 matrix will be about 9 times larger than for the constant kernel solution. 
 Some situations might require order 2 or 3 or even more. Order 3 requires 
 roughly 100 times more calculation than the constant kernel solution. We
 see that the problem becomes quickly UN-tractable, and that one of the
 main advantages of the OIS, the fast computing time will be completely lost.
 Fortunately, it will be demonstrated in this article that the fit of the
 spatial variations of the kernel can be achieved for little additional
 computing cost, provided some reanalysis of the problem. 
\section{Basic equations.}
Here we remind briefly the basics of image subtraction (Alard \& Lupton 1998).
 The essence of the method is to find a convolution kernel ($K$), that will
 transform a reference image ($R$) to fit a given image $I$.\\
In terms of least-squares this is equivalent to look for a kernel
 that will minimize the sum:
\begin{equation}
 \sum_i \left ([R \otimes K](x_i,y_i) - I(x_i,y_i) \right )^2
\end{equation}
 Provided the kernel is decomposed on basis of function, the above 
 equation become a simple linear least-square problem. For the kernel
 decomposition we take: 
$$
   K(u,v) = \sum_n a_n \  K_n(u,v)
$$
 with:
$$
 K_n(u,v) = e^{-(u^2+v^2)} \ u^i \ v^j 
$$
The kernel solution can be calculated by solving the following linear system:
\begin{equation}
  M \bf a = \bf B
\end{equation}
With:
$$ 
 M_{ij} = \int [R \otimes K_i](x,y) \  \frac{[R \otimes K_j](x,y)}{\sigma(x,y)^2} \ dxdy
$$
And:
$$
 B_i = \int I(x,y) \ \frac{[R \otimes K_i](x,y)}{\sigma(x,y)^2} \ dxdy
$$
Additional linear parameters can be included to fit the differential
 background variations between the images. For details see
 Alard \& Lupton (1998).
\section{Solving for non-constant kernel solution with minimum computing time.}
Most of the computing time involved in the computation of the kernel solution
 is spend in the calculation of the least-square matrix. The solution
 of the linear system itself takes an almost negligible amount of time.
 The computing time to build the matrix goes like
 the square of the number of coefficients. Thus the problem is that if we 
 introduce new coefficients in order to fit the kernel variations, the cost
 of the calculation will become quickly prohibitive. To be more specific, 
  let's derive analytical formula's for the spatial variations of the kernel.
  At each position (x,y) we can develop the kernel on the basis of function {$K_n$}.
 Thus the coefficients $a_n$ will be function of (x,y). Consequently the
 kernel can be written:
$$
  K(u,v) = \sum_n a_n(x,y) \  K_n(u,v) 
$$
 We assume that the coefficients $a_n$ are smooth function of x and y.
 For simplicity we adopt a polynomial function of degree $d_1$.
$$
  a_n(x,y) = \sum_{i,j} b_{i,j} \ x^i \ y^j
$$
Let's calculate the relevant Least-square vectors:
$$
 V_n(x,y) =  \ W_m(x,y)  \times P_q(x,y)
$$ 
With:
$$
 W_m(x,y) = \frac{[R \otimes K_m](x,y)}{\sigma(x,y)}
$$
$$
 P_q(x,y) = x^i \ y^j
$$
And:
$$
 q = i+j \times d_1
$$ 
$$
 nc_1 = \frac{(d_1+1)(d_1+2)}{2}
$$
$$
 n = q\ + \ m\times nc_1
$$
The elements of the least-square matrix can be expressed as scalar products
 of the Least-square vectors:
$$
 M_{n1,n2} = \int W_{m1}(x,y) \ W_{m2}(x,y) \ P_{q1}(x,y) \ P_{q2}(x,y) \ dx dy
$$
The previous integral extend over all the image. However, we can always imagine
 to split the integration domain into small rectangular sub-areas. Within
 such small areas, one can always assume that the kernel is constant. It 
 corresponds to approximate locally x, and y by the coordinates of 
 the centers of the sub-areas, ($x_k$, $y_k$). \\
 Considering this approximation the matrix elements can be written: 
\begin{equation}
 M_{n1,n2} = \sum_k Q_{m1,m2}^k(x,y) \  P_{q1}(x_k,y_k) \ P_{q2}(x_k,y_k)
\end{equation}
 With the following definition for the integral in the rectangular domain $D_k$:
 $$
 Q_{m1,m2}^k = \int_{D_k} W_{m1}(x,y) \ W_{m2}(x,y) \ dx dy
 $$
 A similar method can be applied to expand the vector on the right hand side of eq. 2.
\begin{equation}
  B_n = \sum_k \ W_m^k(x,y) P_{q1}(x_k,y_k)
\end{equation}
 With:
 $$
  W_m^k(x,y)= \int_{D_k} I(x,y) \ \frac{W_{m1}(x,y)}{\sigma(x,y)} \ dxdy
 $$
 We see that the matrix elements corresponding to fitting the spatial
 variations of the kernel can be deduced the matrix elements $Q_{m1,m2}^k$
 for only the cost of a summation over the number of sub-areas. The matrix
 $Q_{m1,m2}^k$, corresponds to the constant kernel solution, and has a much
 smaller number of elements than the matrix $M_{n1,n2}^k$. Once $Q_{m1,m2}^k$ 
 has been estimated,  $M_{n1,n2}$ can be calculated using the expansion
 given in Eq. 3. This procedure results in a drastic economy in computing
 time, since the summation inside the individual sub-areas are avoided
 for the calculation of the matrix elements. Using Eq. 4 the same type 
 of expansion applies also to the calculation of the vector $B_n$.
 Since the individual sub-areas
 should be slightly larger than the kernel, a typical number for 
 the surface of a sub-area is $30\times30$ pixels. It means that in computing
 application we save a factor of about 1000 in the calculation of the
 matrix elements of the matrix $M_{n1,n2}$ which corresponds to the
 spatial variations. In practice, fitting the kernel variations costs only
 about 20 \% more than making a constant kernel fit.\\\\
 In high latitude fields, the only useful areas are those centered on
 bright (high S/N) objects. It usually results in a drastic reduction in 
 the number of pixels to fit. Even in crowded fields, most of the information
 is contained in some high density regions. Thus in practise the fit can be
 restricted to a fraction of the image. \\\\
 Of course, differential fitting of the background variations between the 
 images can be included, exactly in the same way as was done for constant
 kernel solutions. 
\section{numerical simulations.}
\subsection{Test with variable PSF.}
\subsubsection{The simulated images.}
 To check the ability of the method to reconstruct the spatial variations
 of the kernel, a serie of test are performed with simulated images. The images
  are generated by putting
 stars randomly in the image, with a magnitude distribution corresponding to a
 bulge luminosity function. Noise is added in the images according to Poisson
 statistics. For the reference image we take a constant gaussian PSF:
 $$
  \Phi_{0}(u,v) = e^{-\alpha(u^2+v^2)}
 $$
 In our simulation we take $\alpha=0.5$. \\
 For the other image we simulate a variable PSF
 by taking the sum of 2 gaussians with different width and a relative weighting
 which is function of the position in the image. We also introduce a 
 normalization factor for conservation of the flux. 
$$
 \Phi(u,v) = \frac{e^{-\alpha(u^2+v^2)} + r(x,y) \times e^{-\frac{\alpha(u^2+v^2)}{4}} }{1+4 \ r(x,y)}
$$
 For the spatial variation we take:
$$
 r(x,y) = \frac{0.25 \ y}{w}
$$
Where $w$ is the size of the simulated image. \\
The resulting images are presented in Fig. 1. 
\begin{figure*} 
\centerline{\psfig{angle=180,figure=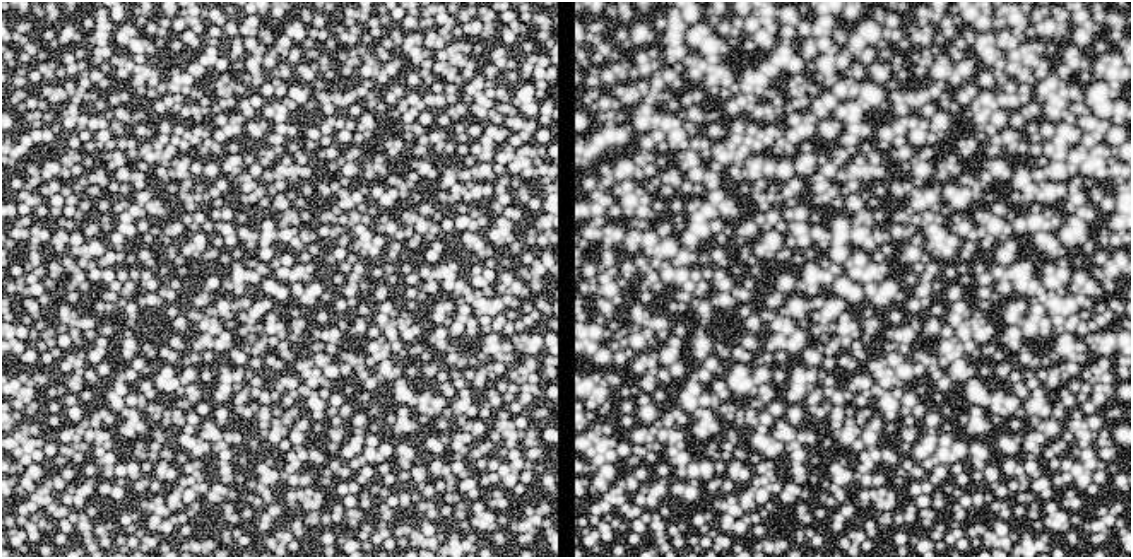,width=15cm}}
\caption{
 Simulation of crowded fields images. Left is the image with constant PSF,
 and next is the image with PSF variations along the Y axis.
 Note the large amplitude of the PSF variations.
 A total of 2500 stars has been included in this simulation. 
  }
\end{figure*}
\subsubsection{Image subtraction.}
 First, a subtracted image with constant kernel solution has been produced.
 We use a set of parameter for the kernel that is similar to Alard \& Lupton 
 (1998), except that one basis function a delta function has been added to
 the kernel representation. Using a delta function might be useful in case
 the image have very similar seeing (Wozniak 1998). For the fit with with
 variable kernel solution we used a polynomial of order 2, since 
 the variations of the coefficients of the PSF function are nearly
 parabolic. The resulting subtracted images are presented in Fig. 2.
\begin{figure*} 
\centerline{\psfig{angle=180,figure=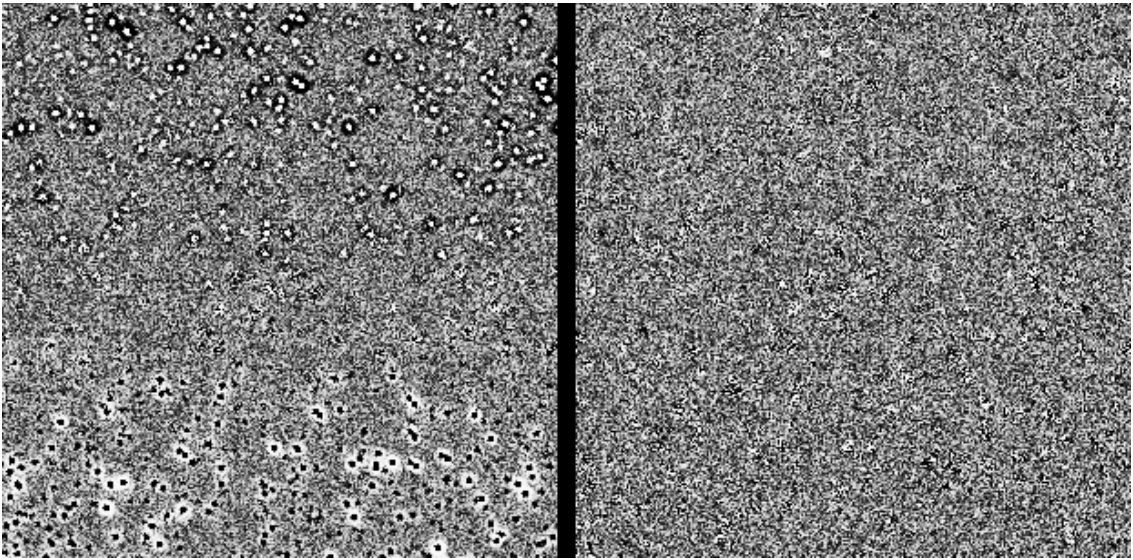,width=15cm}}
\caption{
  Left is the subtracted image obtained with constant kernel solution.
  Note the systematic pattern along the Y axis due to the kernel
  variations. Just next we present the subtracted image obtained
  by fitting the spatial variations of the kernel to order 2.
 }
\end{figure*}
\subsubsection{Noise estimation.}
 Noise in the subtracted image has 2 origins, the noise in the image to fit,
 and the noise in the reference image convolved with the kernel 
 solution .
 In our images the Poisson noise can be extremely well approximated locally 
 by a gaussian distribution with $\sigma=\sqrt{N}$. However this is not true
 for the reference image, since it has been transformed by convolution.
 The noise in the convolved image can be estimated in a straightforward
 way. The convolution will result in the combination of different Gaussian
 distribution, with differents $\sigma$ and weights. We keep the same
 notations, $I$ is the image to fit $R$ is the reference image, and we
 define the $IC$ which is the reference image convolved with the kernel 
 solution.
$$
  IC_i = \sum_j R_{i-j} \ K_j 
$$
 The combination of
 gaussian distribution will result in a gaussian distribution, and the
 resulting $\sigma$ of the distribution can be estimated by calculating 
 the variance:
$$
 \sigma_i^2 = \sum_j var(R_{i-j}) \ K_j^2 = \sum_j R_{i-j} \ K_j^2  
$$
Thus we see that the local variance of the image can be estimated by convolving
 the image with a filter that is just the square of the initial convolution
 filter. Consequently what we call {\it Poisson deviation} will be defined as:
$$
 \delta = \sqrt{\sigma_i^2 + I_i}
$$
The histograms of the pixels in the subtracted images normalized by the 
 Poisson deviation are presented in Fig. 3. It is interesting to note that
 order 2 is sufficient to produce a Chi-square per degree of freedom (Chi2/Dof)
 which is extremely close to 1.
\begin{figure}
\centerline{\psfig{angle=0,figure=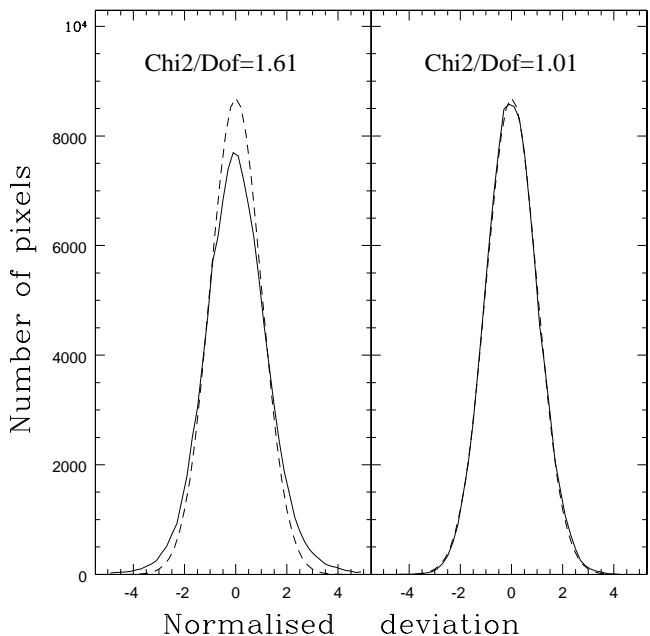,width=9cm}}
\caption{
 Histograms of the normalized deviations in the subtracted images presented in Fig. 2. 
 Left is the
 histogram for constant kernel solution, and right is the histogram for
 a fit of kernel variation to order 2. 
 The pixels in the subtracted images have been normalized by the 
 Poisson deviations (see text for details). The dashed curve is gaussian
 with $\sigma=1$.
 }
\end{figure}
\subsection{Checking the ability of the method to correct the astrometric registration.}
 Image registration is performed by calculating a polynomial transform from the
 positions of brights objects. However, there is no  guarantee that this
 procedure is optimal. 
 In case the image registration to the reference frame is not perfect, 
 a simple translation can be taken into account by a constant kernel solution.
 However more complex features like differential rotation, or
 differential distortion between the images cannot be corrected 
 with constant kernel solution. But they can be corrected with non-constant
 kernel solution. We will illustrate this fact by simulating differential 
 rotation between 2 images. We keep the reference image we had already 
 generated for the previous simulations, and make another one by rotating the
 frames with respect to its center with an amplitude of 0.7 pixel from one
 corner to the other corner of the frame. The result of subtraction are 
 presented in Fig. 4. The systematic pattern due to rotation appears clearly
 in the image with constant kernel solution, while it is completely removed
 in the fit of a solution with a spatial variation of order 2. This is well
 confirmed by the Chi-square analysis which shows that an optimal result
 has been reached with the non-constant kernel solution (see Fig. 5).
\begin{figure*} 
\centerline{\psfig{angle=180,figure=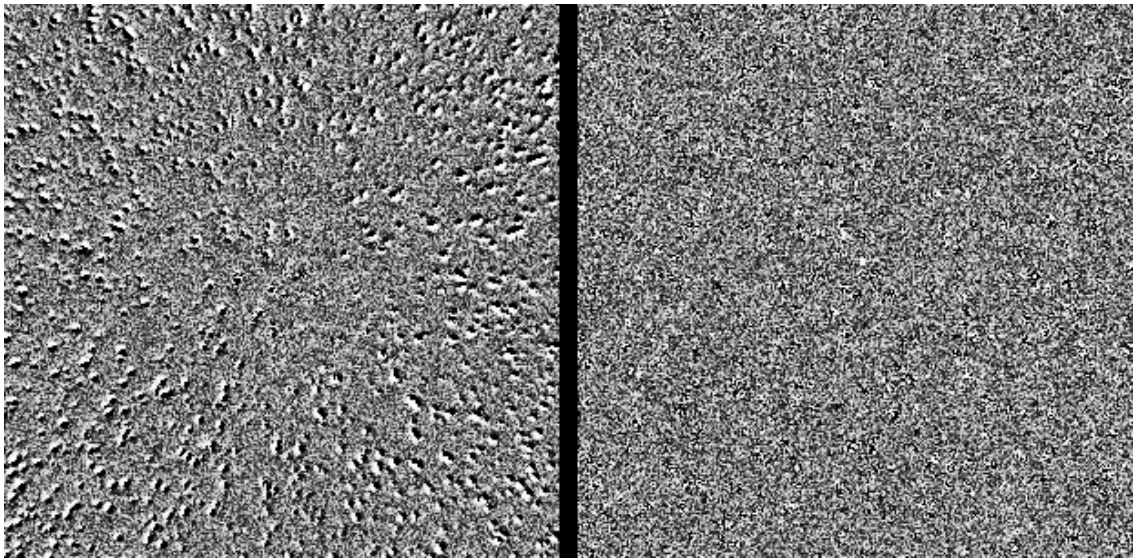,width=15cm}}
\caption{
  Image subtraction in case of images mis-aligned by differential rotation.
  Left is the subtracted image obtained with constant kernel solution.
  Note the systematic pattern due to the differential rotation between
  the images. Next is the subtracted image obtained
  by fitting the spatial variations of the kernel to order 2. Note the
  complete diseaperance of the systematic patterns. 
 }
\end{figure*}
\begin{figure} 
\centerline{\psfig{angle=0,figure=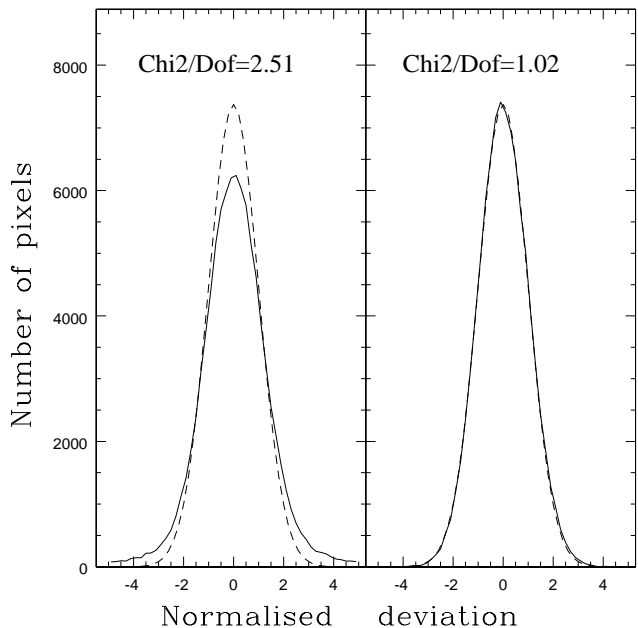,width=9cm}}
\caption{
 Histograms of the normalized deviations in the subtracted images presented in Fig. 4. 
 Left is the histogram for constant kernel solution, and right is the histogram for
 a fit of kernel variation to order 2. The dashed curve is gaussian
 with $\sigma=1$. Note the goodness and the dramatic reduction of the
 Chi-square when fitting the kernel variations.
 }
\end{figure}
\subsection{Under-sampling.}
 One last problem that can be encountered in astronomical images is 
 under-sampling. To test the sensitivity of the method to under-sampling,
 we simulate a pair of very under-sampled images. We take $\alpha=2.0$
 ($FWHM = 1.17$ pixels) for the reference and $\alpha=1.0$ 
 ($FWHM = 1.67$ pixels) for the other image. Since our goal is just to test 
 the effect of under-sampling only, we perform image subtraction with
 constant kernel solution. The resulting normalized Chi-square
 distribution is presented in Fig. 6. The result is as good as in previous
 simulations. One may wonder why under-sampling does not induce any problem,
 since under-sampling should affect the convolution with the basis vectors.
 It is true that individually the least-square vectors which are obtained with
 under-sampled images will differ from the well sampled vectors. However one
 may not forget that the best solution is constructed by taking linear
 combinations of the least-square vectors. Even if the vectors are slightly
 different from the well sample vectors, an optimal linear combination of
 these vectors can be made anyway.
\begin{figure} 
\centerline{\psfig{angle=0,figure=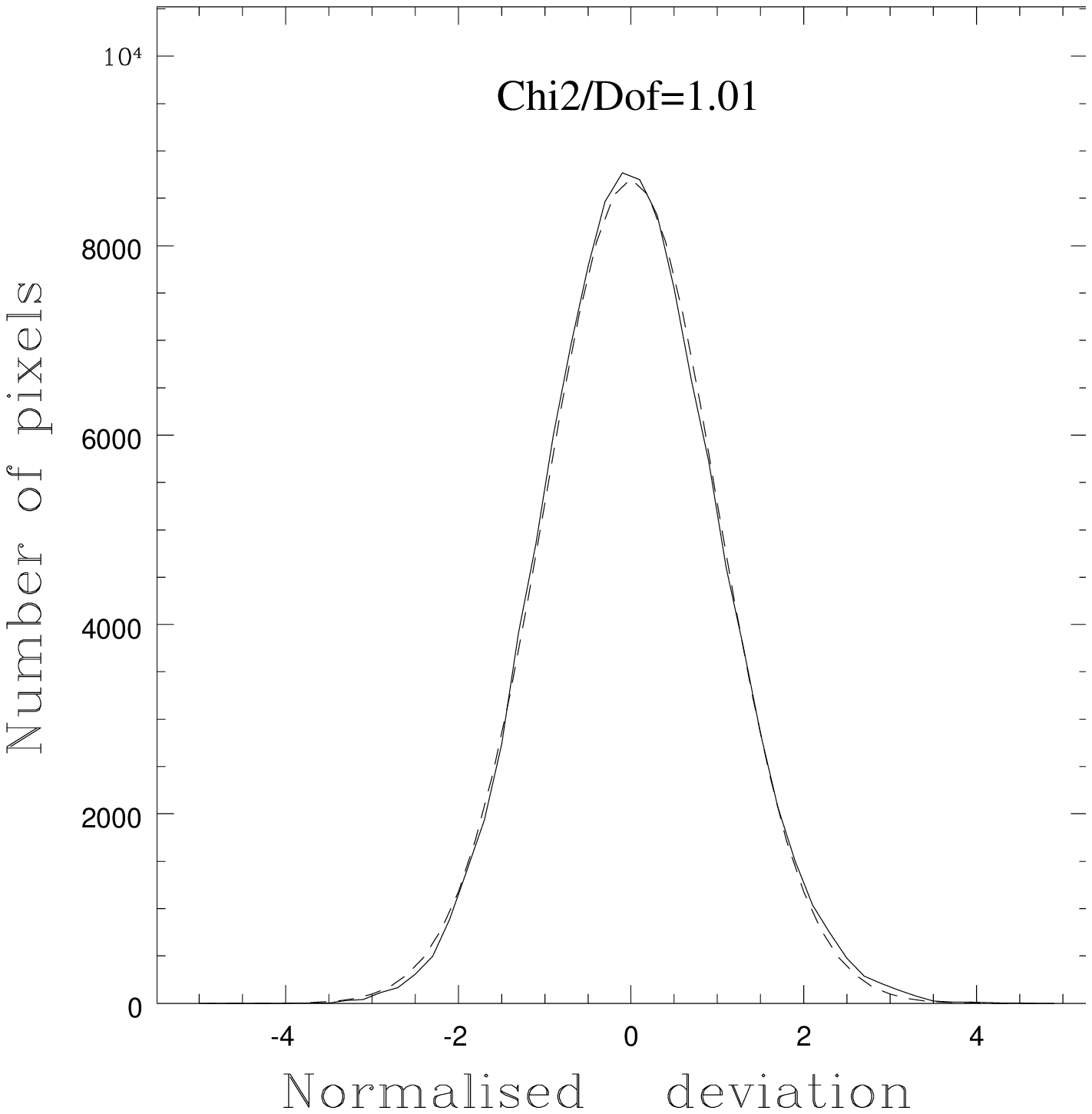,width=8cm}}
\caption{
 Histograms of the deviations in the subtracted image for simulated,
 very under-sampled images. The PSF
 in the reference image is very narrow with: $FWHM = 1.17$ pixels, thus
 we are in presence of a bad case of under-sampling. For the other
 image  $FWHM = 1.67$, consequently it is also under-sampled.
 }
\end{figure}
\section{Testing with astronomical images.}
It was already shown in Alard \& Lupton (1998) that nearly optimal
 results (Chi2/Dof = 1.05) could be achieved within a small sub-area
 of a crowded field. This is of course not true if the area
 is extended, since kernel variations are not negligeable any more.
 However, by using a variable kernel solution, one should be able
 to get a result closely similar for a larger region to what had been
 obtained for a small area. We test this assumption by extracting 2 larger
 sub-areas ($256 \times 256$) from the fields already used in Alard \&
 Lupton (1998), and making image subtraction with constant and variable
 kernel solutions. The results are shown in Fig. 7 and Fig. 8. The constant
 kernel solution achieved only Chi2/Dof = 1.19, with numerous systematic 
 residuals near the edges of the field. On the contrary the subtracted image
 achieved with variable (Chi2/Dof = 1.04) is very close to the Chi-square 
 obtained with constant kernel solution for a smaller image. This analysis
 demonstrates the ability of the method to deal with kernel variations for
 crowded field images. It is certainly useful in this case, since larger areas
 and thus slightly more robust and reliable results can be obtained in crowded
 fields. But of course the ability to deal with kernel variations is absolutely
 essential when one has to deal with fields having low density of bright
 objects. This is the case of supernovae search, Cepheids surveys in other
 galaxies, and also in the case of monitoring of gravitational mirages. 
 Thus the method has many important applications, and an illustration can
 be found in the analysis of a serie of images of the Huchra Lens gravitational
 mirage (Alard \&Wozniak 1998).
\begin{figure*}
\centerline{\psfig{angle=180,figure=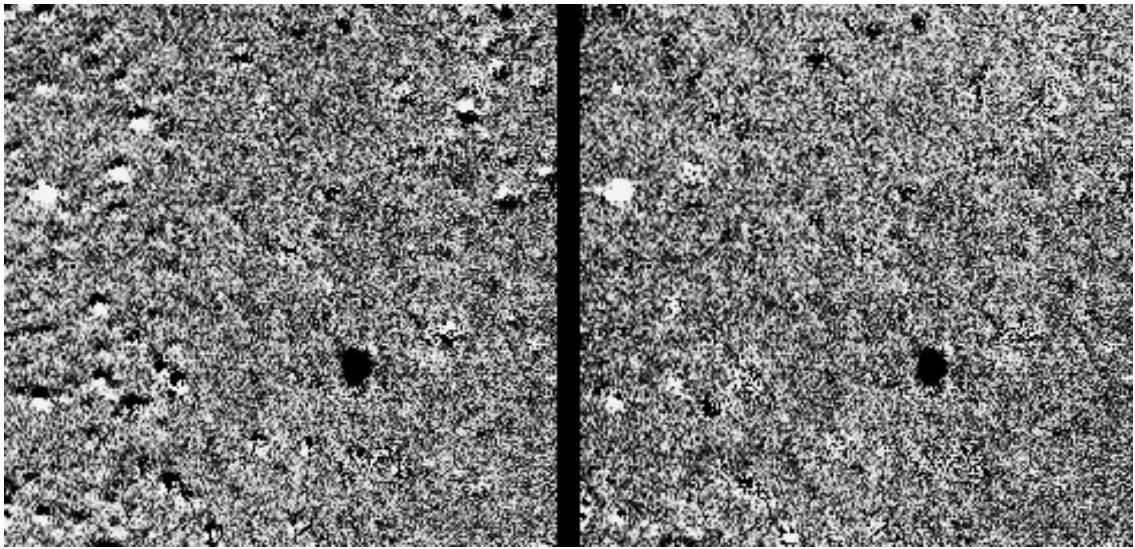,width=15cm}}
\caption{
 Image subtraction using OGLE II exposures of Galactic Bulge fields.
 Left is the subtracted image obtained with constant kernel solution.
  Note the systematic residuals around the bright objects on the left
 side of the image and in the upper right corner.
 Note the diseaperance of these pattern in the next image, which has been 
 obtained by fitting the spatial variations of the kernel to order 2.
 Two variables are present in the field (the bright and dark spots)
 , and circular areas around these objects had to be excluded for Chi-square
 evaluation.  
 }
\end{figure*}
\begin{figure}
\centerline{\psfig{angle=0,figure=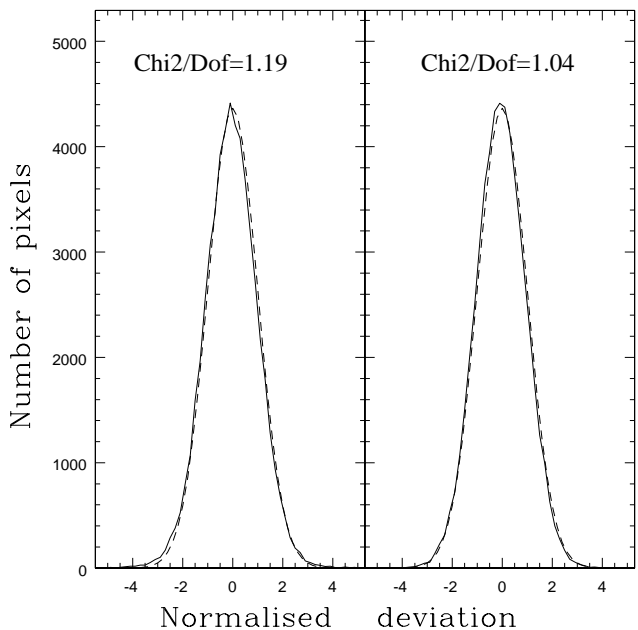,width=9cm}}
\caption{
 Histograms of the normalized deviations in the subtracted images presented in Fig. 7
 (OGLE II data). Left is the
 histogram for constant kernel solution, and right is the histogram for
 a fit of kernel variation to order 2. The dashed curve is gaussian
 with $\sigma=1$.
 }
\end{figure}
\section{Conclusion and summary.}
It has been demonstrated that image subtraction with non-constant kernel solution
 could be achieved in minimum computing time. The cost to perform non-constant
 solution is only 20 \% to 30 \% in excess of the constant kernel solution. 
 The ability of the method to deal with PSF variations between the images
 is demonstrated using Monte-Carlo simulation of stellar fields. It is found that
 even large relative PSF variations are very well corrected by the method, resulting
 in a Chi2/Dof very close to 1 in the subtracted image. Non constant kernel solution
 can also automatically correct for imperfect registration between the images. This
 possibility is illustrated by generating 2 images with differential rotation with
 respect to each other. While the systematic pattern due to differential rotation
 is visible in the subtracted image obtained with constant kernel solution, it 
 disappear completely by fitting the spatial variations of the kernel. To complete
 the analysis with Monte-Carlo simulation, we tackle the problem of under-sampling
 by generating 2 very under-sampled images of the same stellar field. A chi-square
 analysis shows that even in this case an optimal result can be achieved. Finally
 we apply the method to real astronomical data. It is shown that nearly optimal
 results can be achieved with non-constant solution, even if the size of the sub-area
 used for the fit is increased. This result is certainly useful for crowded fields data,
 but proved to be completely essential in case of fields with low density of high
 signal to noise objects. A good example is the case of the photometry of the 4 images
 of the Huchra lens (Wozniak \& Alard 1999) To conclude, we can say that new extension
 of the image subtraction method will certainly prove to be most useful for the 
 supernovae search that are currently undertaken, but may also prove useful for 
 variability search in other galaxies, survey of variables near the core of globular
 clusters (Olech {\it et al.} 1999), and may also become one of the favorite method to analyze
 microlensing variability in gravitational lens systems.\\\\\\
{\bf Software availability: \\}
A set of C programs to perform image subtraction can be obtained on request
 to C. Alard (alard@iap.fr). Further documentations and C packages will 
 be soon available on a public access Web site.
\section{Acknowledgments}
 It is a pleasure to thank  B. Paczy\'nski for his hospitality during my
 stay in Princeton. I would like to acknowledge support from
 NSF grant AST 95-30478. I thank B. Paczynski, P. Wozniak, and R. Lupton 
 for many interesting discussions.

%
\end{document}